\documentclass[aps,prb,twocolumn,showkeys,floatfix]{revtex4}

\usepackage{graphicx,color}

\newcommand{\rev}[1]{\textcolor{black}{#1}}

\graphicspath{{figs/}}
\begin{document}
\title{Auxetic Nanomaterials:  Recent Progress and Future Development}
\author{Jin-Wu Jiang}
    \altaffiliation{Email address: jiangjinwu@shu.edu.cn; jwjiang5918@hotmail.com}
    \affiliation{Shanghai Institute of Applied Mathematics and Mechanics, Shanghai Key Laboratory of Mechanics in Energy Engineering, Shanghai University, Shanghai 200072, People's Republic of China}

\author{Sung Youb Kim}
    \affiliation{Department of Mechanical Engineering, Ulsan National Institute of Science and Technology, Ulsan 44919, South Korea}

\author{Harold S. Park}
    \affiliation{Department of Mechanical Engineering, Boston University, Boston, Massachusetts 02215, USA}

\date{\today}
\begin{abstract}

Auxetic materials (materials with negative Poisson's ratio) and nanomaterials have independently been for many years two of the most active research fields in material science.  Recently, these formerly independent fields have begun to intersect in new and interesting ways due to the recent discovery of auxeticity in nanomaterials like graphene, metal nanoplates, black phosphorus, and others.  Here we review the research emerging at the intersection of auxeticity and nanomaterials. We first survey the atomistic mechanisms, both intrinsic and extrinsic, that have been found, primarily through atomistic simulations, to cause auxeticity in nanomaterials.  We then outline the available experimental evidence for auxetic nanomaterials.  In order to lay the groundwork for future work in this exciting area, we close by discussing several future prospects as well as the current challenges in this field.

\end{abstract}

\maketitle
\tableofcontents

\section{Introduction}

The Poisson's ratio, \rev{$\nu_{xy}=-\frac{\epsilon_y}{\epsilon_x}$}, characterizes the resultant strain in the y-direction for a material under longitudinal deformation in the x-direction. The Poisson's ratio is typically a positive number, and has a value around 0.3 for many engineering materials (e.g. steels). The value is positive when a material contracts in the transverse directions when stretched uniaxially. In the uniconstant elasticity theory,\cite{WeinerJH1983} atoms are treated as point particles in a centrosymmetric lattice with only longitudinal interactions. The tensorial elastic constants of the anisotropic solid are related by the Cauchy relations, while the Cauchy relations yield a constant value of 1/4 for the Poisson's ratio in isotropic solids.

However, uniconstant elasticity theory has not been used for many decades, one reason being that it was subsequently found that the Poisson's ratio is not a constant value of 1/4 for all materials.  Instead, classical elasticity theory, which accounts for both longitudinal and transverse interactions\cite{LandauLD}, was found to better represent the Poisson effect and Poisson's ratio in solids. There are two independent parameters in the classical elasticity theory; i.e., the Lam$\acute{e}$ coefficients $\lambda$ and $\mu$, or the bulk modulus $K=\lambda+\frac{2\mu}{3}$ and the shear modulus $\mu$. Instead of a constant value, the Poisson's ratio in classical elasticity theory depends on the ratio between the bulk modulus and the shear modulus, e.g. $\nu=\frac{1}{2}(1-1/(\frac{K}{\mu} + \frac{1}{3}))$ for three-dimensional isotropic materials. The Poisson's ratio is limited to the range $-1< \nu < 0.5$ for three-dimensional isotropic materials within classical elasticity theory.

Within classical elasticity theory, materials are thus allowed to exhibit a negative Poisson's ratio (NPR), which are also known as auxetic materials.\cite{EvansKE1991Endeavour} One way in which the impact of NPR can be gleaned is to note that there exist certain physical properties that are inversely proportional to $1+\nu$ or $1-\nu^2$, which implies that these properties become infinitely large in the limit of the Poisson's ratio $\nu\rightarrow -1$. For example, the \rev{speed of sound} is proportional to $(1+\nu)^{-1/2}$, and the material hardness is related to $(1-\nu^2)^{\gamma}$, with $\gamma$ as a constant. Hence, materials with NPR typically have novel properties such as enhanced toughness and enhanced sound and vibration absorption.

In 1987, Lakes performed seminal experiments to illustrate the NPR in a foam structure.\cite{LakesRS1987sci} Since then, many researchers have demonstrated that the NPR phenomenon is actually quite common both as an intrinsic material property (i.e., NPR occurs without any external engineering of the material structure or composition.) and also in engineered structures.\cite{RothenburgL1991nat,LakesR1993adm,BaughmanRH1993nat,EvansKE2000adm,YangW2004jmsci,RaviralaN2007jms,LethbridgeZAD2010am,BertoldiK2010am,AldersonK2012pssb,clausenAM2015}  For example, the Poisson's ratio was found to be anisotropic in some cubic elemental metals. While the Poisson's ratio is positive along the axial directions in the cubic elemental metals, 69\% of the cubic elemental metals have intrinsic NPR along a non-axial direction.\cite{MilsteinF1979prb,BaughmanRH1998nat} A more recent work has found that the Poisson's ratio for FCC metals can be negative along some principal directions by proper control over the transverse loading.\cite{hoPSSB2016b}

Concurrently, nanomaterials, encompassing such well-known materials like buckyballs, carbon nanotubes, graphene, nanowires, black phosphorus, MoS$_2$ and others, have drawn significant interest within the past two decades.  Within the last three years, the auxetic property has been found in some of these nanomaterials, with the mechanisms underlying the auxetic properties often being due to specific nanoscale physical properties.  Some of these new findings were mentioned in a recent review on auxeticity by Huang and Chen,\cite{HuangC2016am} but a comprehensive review on this emerging field of auxetic nanomaterials is still lacking.  Our objective in this review is to survey the novel mechanisms underpinning auxetic behavior in nanomaterials, and to discuss challenges and opportunities for future work.  We do not discuss auxeticity in bulk materials, for which readers are referred to previous review articles.\cite{LakesR1993am,ChanN1997jms,EvansKE2000adm,EvansKE2000ese,YangW2004jmsci,LiuQ2006,AldersonA2007jae,LiuY2010sre,GreavesGN2011nm,PrawotoY2012cms,HuangC2016am}

\section{Auxetic mechanisms for nanomaterials}

We now discuss the mechanisms that enable the emergence of auxeticity in nanomaterials.  The mechanisms can be delineated as intrinsic, and extrinsic, with the intrinsic mechanisms discussed first. Again, we emphasize that intrinsic mechanisms are those that cause NPR in the material without any external engineering of the material structure or composition.

\subsection{Intrinsic}

\subsubsection{Puckered Crystal Structure}

\begin{figure}[tb]
  \begin{center}
    \scalebox{1.0}[1.0]{\includegraphics[width=8cm]{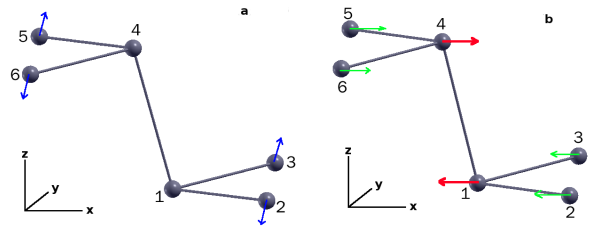}}
  \end{center}
  \caption{(Color online) The evolution of local structure in single-layer \rev{black phosphorus} during uniaxial tension in the y-direction. (a) \rev{Black phosphorus} is stretched in the y-direction, i.e atoms are moved in the direction of the attached arrows (blue online). (b) To accommodate the tension in the y-direction, \rev{black phosphorus} contracts in the x-direction, i.e atoms 1 and 4 move inward along the attached arrows (red online). The 1-4 bond thus becomes more closely aligned with the vertical (z)-direction.  The green arrows display the movement of the four surrounding atoms following the movement of atoms 1 and 4. Reprinted by permission from Macmillan Publishers Ltd: Nature Communications\cite{JiangJW2014bpnpr}, copyright 2014.}
  \label{fig_cfg_poisson}
\end{figure}

\textbf{Black phosphorus.} Black phosphorus is one of the recent entries to the 2D materials canon, which has drawn attention for its potential as an alternate electronic material to graphene\cite{LiL2014,LiuH2014,BuscemaM2014}.  It is characterized by its puckered atomic structure, where Fig.~\ref{fig_cfg_poisson} shows the smallest puckered cell. There are two groups of atoms, with 4, 5, and 6 in the top group and 1, 2, and 3 in the bottom group. This puckered structure can be \rev{conceptually} obtained geometrically as follows:  assuming both top and bottom atoms are initially in a planar honeycomb lattice in the xy plane, compression of the planar lattice in the x-direction will result in puckering of the structure into the top and bottom groups.

This puckered structure is highly anisotorpic. More specifically, this puckered structure is elastically softer in the x-direction, owing to the construction of inter-group angles like $\theta_{146}$, so the in-plane Poisson's ratio $\nu_{yx}$ is large. As a direct result of the anisotropic puckered structure, the Poisson's ratio in the z-direction is negative, i.e. the thickness in the z-direction increases during the deformation of the black phosphorus along the y-direction.\cite{JiangJW2014bpnpr} This occurs because when the structure is stretched in the y-direction, it undergoes a large contraction along the x-direction due to the large value of $\nu_{yx}$, leading to the decrease of inter-group angles like $\theta_{146}$. That is, the inter-group bond 1-4 will be aligned closer to the z-axis, which results in the expansion of the thickness in the z-direction. Interestingly, the pucker can also be regarded as two coupling hinges formed by the angles $\theta_{546}$ and $\theta_{214}$, which leads to a nanoscale version of the coupling hinge mechanism. The NPR is thus closely related to the condition of $\theta_{146}>90^{\circ}$ in black phosphorus. It should be noted that the out-of-plane NPR exists concurrently with a large positive value of the in-plane Poisson's ratio $\nu_{yx}$.  

While the NPR in the z-direction of black phosphorus was discussed in detail in Ref.~\onlinecite{JiangJW2014bpnpr,JiangJW2014bpsw}, this effect has also been mentioned in some other works, which applied mechanical strains to black phosphorus. For example, the NPR was also observed during the investigation of strain effects on the electronic properties of single-layer black phosphorus\cite{ElahiM2014prb} or the thermoelectric properties for bulk black phosphorus.\cite{QinGarxiv14060261}

\textbf{Other puckered nanomaterials.} Orthorhombic arsenic shares the same puckered atomic configuration as black phosphorus, and thus is expected to exhibit NPR.  This was confirmed in recent first principles calculations by Han et al. for few-layer orthorhombic arsenic.\cite{HanJ2015ape} The Poisson's ratio was found to be about -0.093 for single-layer orthorhombic arsenic, while becoming increasing negative with increasing numbers of layers.  
The saturation value for the NPR, about -0.13, is reached for the few-layer orthorhombic arsenic with layer number above four. We close this discussion by noting that the unfolding of the puckered structure is a mechanism that is operant at many length scales, from the nanoscale as seen here to the macroscale, as shown previously.\cite{LakesRS1987sci,FriisEA1988jms}

\subsubsection{Competition between deformation modes}

While some nanomaterials like black phosphorus exhibit NPR for all strains due to their crystal structure, other nanomaterials exhibit intrinsic NPR due to other mechanisms. One is related to the basic deformation mechanisms that nanomaterials undergo, which make either a positive or negative contribution to the Poisson's ratio.  Due to this balance, nanomaterials can exhibit intrinsic NPR when the mechanisms leading to NPR dominate, as explained in detail below.

\textbf{Possible auxeticity in single-walled carbon nanotubes:} Analytic expressions for the Poisson's ratio are useful for studying mechanisms by which auxeticity can be induced in nanomaterials.  For achiral (armchair or zigzag) single-walled carbon nanotubes, analytic expressions were reported in several works.  In 2003, Chang and Gao derived the analytic formula for the Poisson's ratio (and Young's modulus) for achiral single-walled carbon nanotubes, using bond stretching and angle bending potentials.\cite{ChangT2003jmps} With a similar molecular mechanics approach, Shen and Li also obtained the analytic formula for the Poisson's ratio (and Young's modulus) of achiral single-walled carbon nanotubes in 2004.\cite{ShenL2004prb} Chang et al. also derived the Poisson's ratio (and Young's modulus) for single-walled carbon nanotubes of arbitrary chirality in 2005.\cite{ChangTC2005apl} Wu et al. used the molecular mechanics approach to derive the shear modulus, Poisson's ratio and Young's modulus for achiral single-walled carbon nanotubes in 2006.\cite{WuY2006tws}

In 2008, Yao et al. generalized the above analytic expressions for the Poisson's ratio to allow the difference between two inequivalent C-C bond lengths in achiral single-walled carbon nanotubes.\cite{YaoYT2008pssb} The following generalized expressions explicitly show the dependence of the Poisson's ratio on the structural parameters (bond length and angles), and force constants for armchair and zigzag single-walled carbon nanotubes,
\begin{eqnarray}
\nu_{\rm arm} & = & \frac{\cos\frac{\alpha}{2}\left[\frac{b^{2}C_{b}}{4\left(C_{\alpha}+\Delta C_{\beta}\right)}-1\right]}{\left(\frac{a}{b}+\cos\frac{\alpha}{2}\right)\left[\frac{b^{2}C_{b}}{4\left(C_{\alpha}+\delta C_{\beta}\right)\tan^{2}\frac{\alpha}{2}}+1\right]}\\
\nu_{\rm zig} & = & \frac{\cos\beta\left(\frac{a}{b}-\cos\beta\right)\left[1-\frac{b^{2}C_{b}}{2\left(4\delta C_{\alpha}+C_{\beta}\right)}\right]}{\left[\frac{2C_{b}}{C_{a}}+\cos^{2}\beta+\frac{b^{2}C_{b}\sin^{2}\beta}{2\left(4\delta C_{\alpha}+C_{\beta}\right)}\right]},
\end{eqnarray}
where $C_a$ and $C_b$ are force constants for bonds $a$ and $b$, while $C_{\alpha}$ and $C_{\beta}$ are force constants for angles $\alpha$ and $\beta$. Based on these analytic expressions, they performed a speculative examination on the evolution of the Poisson's ratio by varying a single parameter (or ratio of parameters) while leaving other parameters unchanged. It was found that auxeticity in single-walled carbon nanotubes is possible, though under conditions that are physically difficult to realize.  For example, through the relation between the Poisson's ratio and the angle $\alpha$, the appearance of auxeticity requires the $\alpha$ to be larger than $200^{\circ}$, while the actual value of $\alpha$ around $120^{\circ}$.

\begin{figure}[tb]
  \begin{center}
    \scalebox{1.0}[1.0]{\includegraphics[width=8cm]{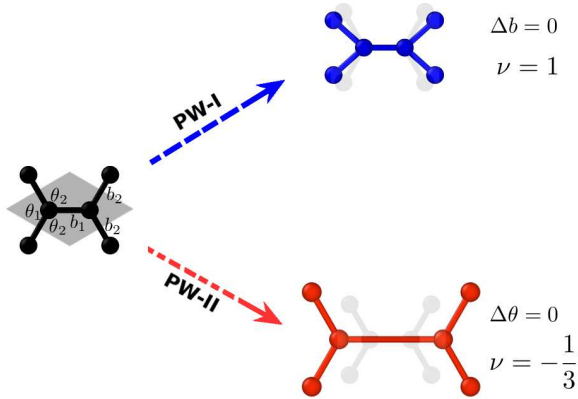}}
  \end{center}
  \caption{(Color online) Two typical ideal deformation pathways during the tensile deformation of graphene. The left atom cluster (black online) is stretched along the horizontal x-direction. The parallelogram gray area indicates the unit cell. PW-I (blue online): carbon-carbon bond lengths remain constant ($\Delta b=0$), while angles are altered to accommodate the external strain, which results in a Poisson's ratio of $\nu=1$. PW-II (red online): angles are unchanged and bond lengths are elongated to accommodate the external tension, resulting in a NPR of $\nu=-1/3$. The lighter shades show the undeformed structure. Reproduced with permission from Ref.~\onlinecite{JiangJW2016npr_intrinsic}. Copyright 2016, American Chemical Society.}
  \label{fig_pathway}
\end{figure}

\begin{figure}[tb]
  \begin{center}
    \scalebox{1.0}[1.0]{\includegraphics[width=8cm]{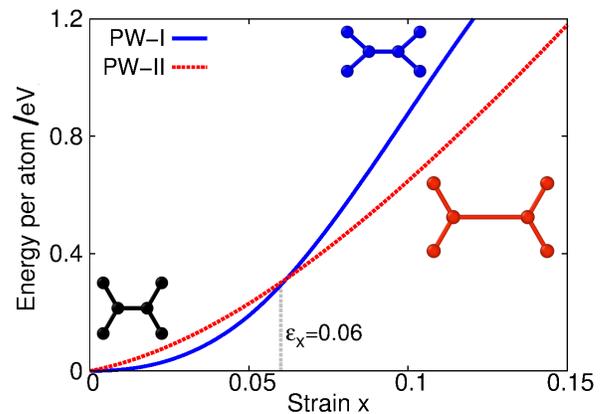}}
  \end{center}
  \caption{(Color online) Pathway energy curve for PW-I and PW-II deformation modes.  The curves show a crossover at $\epsilon_x=0.06$, which predicts a transition from PW-I mode (positive Poisson's ratio) to PW-II mode (negative Poisson's ratio) during the tensile deformation of graphene. Left bottom inset (black online) shows the undeformed structure. Top inset (blue online) displays the PW-I deformed structure. Right inset (red online) is the PW-II deformed structure. Reproduced with permission from Ref.~\onlinecite{JiangJW2016npr_intrinsic}. Copyright 2016, American Chemical Society.}
  \label{fig_bondangle_energy}
\end{figure}

\textbf{Monolayer graphene:} Recently, two of the present authors and their collaborators found that the Poisson's ratio in bulk monolayer graphene is strain dependent, and changes from positive to negative at the critical strain of 6\%, indicating intrinsic auxeticity for monolayer graphene.\cite{JiangJW2016npr_intrinsic} The auxeticity of graphene is intrinsic, because it occurs in pristine graphene without any external modifications to the structure, shape or composition of graphene.  However, the fact that the NPR occurs at a finite, non-zero strain implies that this effect is a highly nonlinear one for graphene.

The Poisson's ratio for graphene has also been derived analytically,\cite{ScarpaF2009nano} or can be obtained directly from the analytic formula for the Poisson's ratio of nanotubes\cite{ChangT2003jmps,ShenL2004prb,WuY2006tws} in the limit of large tube diameters. For example, from Chang and Gao's results, the Poisson's ratio for graphene is,\cite{ChangT2003jmps}
\begin{eqnarray}
\nu = \frac{K_b a^2/K_{\theta} - 6}{K_b a^2/K_{\theta} + 18},
\label{eq_nu_vffm}
\end{eqnarray}
where $a=1.42$~{\AA} is the C-C bond length, $K_{b}$ is the force constant characterizing resistance to stretching, and $K_{\theta}$ is the force constant characterizing resistance to bond angle bending. Eq.~(\ref{eq_nu_vffm}) is applicable for both armchair and zigzag directions in graphene because the Poisson's ratio is isotropic in graphene, as required by the three-fold rotational symmetry in the honeycomb lattice structure.\cite{BornM}  

There are two typical deformation pathways shown in Fig.~\ref{fig_pathway} according to Eq.~(\ref{eq_nu_vffm}).  For the first deformation pathway (PW-I), \rev{bonds are difficult to stretch but the bond angles can readily be altered (i.e. $V_{b}>>V_{\theta}$)}, so mechanical strains will be accommodated by only varying angles during the deformation process. For the second deformation pathway (PW-II), \rev{the angles are difficult to change while bonds can more readily be stretched (i.e. $V_{\theta}>>V_{b}$)}, so bond lengths are altered in response to the applied strain. The PW-I mode leads to a positive Poisson's ratio, while the PW-II mode reduces the Poisson's ratio. Hence, the actual Poisson's ratio for graphene depends on the competition between these two deformation modes.

To characterize the competition between these two deformation modes, an energy \rev{criterion} was proposed based on the above two deformation pathways to determine the sign of the Poisson's ratio in bulk graphene. The criteria states that the tensile deformation process for graphene is governed by the deformation mode with lower pathway energy. The pathway energy is computed based on the potential energy of the structure that is manually deformed according to the PW-I or PW-II mode. Applying this energy \rev{criterion}, Fig.~\ref{fig_bondangle_energy} shows that the PW-I mode will be the dominant deformation mode for graphene for strain less than 6\%, while the PW-II mode will dominate the deformation of graphene for strain larger than 6\%. As a result, the Poisson's ratio in graphene with change its sign at the critical strain around 6\%, which explains the numerical simulation results. The pathway energy \rev{criterion} can readily be extended to three-dimensional isotropic materials, where stretching and shearing are two distinct  deformation modes with opposite contribution to the Poisson's ratio.

\subsubsection{Surface and Edge Stress Effects}

One of the defining characteristics of nanomaterials is their intrinsically large surface to volume ratio (for nanowires, quantum dots and nanoplates), or equivalently their large edge to area ratio (for 2D materials).  Specifically, the surfaces and edges lead to surface, or edge stresses\cite{cammarataPSS1994,haissRPP2001} which result from the fact that surface and edge atoms have a lower coordination number (number of bonding neighbors) than atoms that lie within the bulk material, and which are intrinsic to nanomaterials.

These surface and edge stresses can play a dominant role on the mechanical behavior at the nanoscale, leading to unique physical properties that are not seen in the corresponding bulk material.  For example, surface stress alone can cause a phase transformation in a FCC nanowire with initial cross-section area below 2~{nm$^2$}.\cite{DiaoJ2003nm}, and also shape memory and pseudoelasticity in FCC metal nanowires.\cite{ParkHS2005prl,LiangW2005nl} The compression of nanowires owing to surface stress induces nonlinear elastic stiffening or softening, depending on the axial loading direction.\cite{LiangH2005prb}  Edge stresses in graphene have been shown to induce rippling and warping for the edges of graphene ribbons.\cite{ShenoyVB}  Furthermore, as we shall now discuss, the Poisson's ratio of metals and graphene can be changed significantly at nanoscale because of surface or edge effects.

\textbf{Surface stress induced auxeticity for metal nanoplates.} For two-dimensional metal nanoplates, free surfaces can strongly influence the mechanical properties, especially in thin nanoplates with thickness of a few nanometers or less. If the surface stress is tensile, as is typical for FCC metals\cite{wanMSMSE1999}, there are induced compressive stresses along the in-plane directions which balance the tensile surface stresses, where the induced compressive stresses are  inversely proportional to the nanoplate thickness.\cite{HoDT2014nc,HoDT2015pssb} One of the present authors and his collaborators found that the compressive induced stresses in many metal nanoplates can lead to auxeticity, even though these metals are not auxetic in their bulk form.\cite{HoDT2014nc}

\begin{figure}[tb]
  \begin{center}
    \scalebox{1.0}[1.0]{\includegraphics[width=8cm]{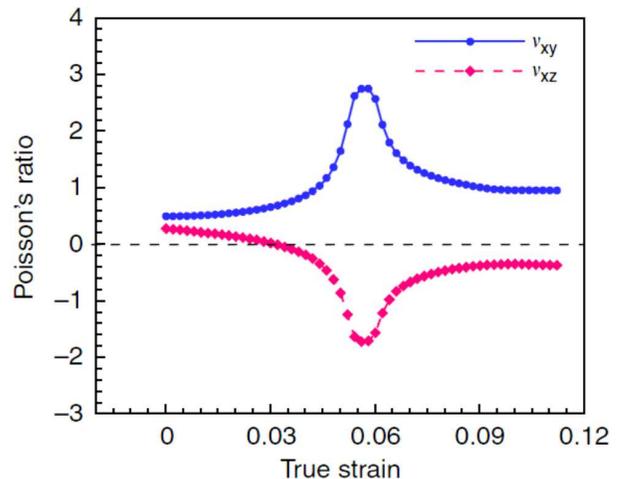}}
  \end{center}
\caption{The Poisson's ratio of an Al (001) nanoplate with a thickness of $5a_0$ (2.02~{nm}) under uniaxial tensile stress along the [100]-direction. Reprinted by permission from Macmillan Publishers Ltd: Nature Communications\cite{HoDT2014nc}, copyright 2014.}
  \label{fig_2014nc_fig1a}
\end{figure}

For illustration, consider Al nanoplates with (100) surfaces, where the Poisson's ratio in the thickness direction becomes strain dependent and can be negative in a particular strain range as illustrated in Fig.~\ref{fig_2014nc_fig1a}, where we note that larger compressive stresses yield more auxetic behavior. The strain dependence of the Poisson's ratio demonstrates that the NPR is a highly nonlinear effect in the nanoplates, which results from the effect of surface stress as well as the loading direction.

The auxeticity of the nanoplates can be increased in several ways. For example, decreasing the thickness of the nanoplate enhances the induced compressive stress. Alternatively, increasing the temperature elastically softens the material, and as a result the effect of the induced compressive stress on the auxeticity is stronger.\cite{HoDT2015pssb} The surface stress-induced auxeticity was found to be common to FCC metals, as it was observed in FCC(001) nanoplates of several metals including Al, Ni, Cu, Pd, Ag, Pt and Au. It is important to mention that surface stresses only induced auxetic behavior for FCC (001) nanoplates under loading along the [100]-direction.  A similar phenomenon was observed in BCC Fe(001) nanoplates under uniaxial compression.

\textbf{Synergetic effects of surface stress and cross-sectional shape for auxetic nanowires.} Inspired by the above works on the NPR of metal nanoplates, Kim and collaborators found that the Poisson's ratios of several nanowires not only depend on their sizes via surface stress but also depend on their shape, i.e., the aspect ratio (width to thickness ratio) of the cross-section.\cite{HoDT2016sr} For square cross sections, the Poisson's ratios $\nu_{xy}$ and $\nu_{xz}$ have the same positive value. For rectangular cross sections with the aspect ratio above 1, both $\nu_{xy}$ and $\nu_{xz}$ are dependent on the applied axial strain. The Poisson's ratio in the thickness direction ($\nu_{xz}$) becomes negative for strains above a critical value (typically between 3\% to 6\%). The Poisson's ratio value is more negative for nanowires of larger width to thickness ratio, and with the aspect ratio of 2.5, the Poisson's ratio is relatively close to that of the corresponding nanoplate with the same thickness.

The auxeticity for the metal nanowires was attributed to the asymmetric surface induced stresses in the width and the thickness directions. The magnitude of the surface induced stress in the width direction is larger than that in the thickness direction, which leads to the increase (decrease) of the Poisson's ratio in the width (thickness) direction. The Poisson's ratio in the thickness direction can be driven to be negative when the difference between these two surface induced stresses is sufficiently large. Here, asymmetric stresses were used to imply the different magnitudes of the two induced stresses along the lateral directions. This asymmetric stress is induced by the free surface effect. One can enhance the auxeticity by introducing a hole at the center of the nanowire cross section, such that the nanowire becomes a nanotube with increased surface area.  It is noteworthy to mention that in the case of the nanoplate, the induced compressive stress along the in-plane lateral direction is non-zero whereas that along the thickness direction is zero, so the mechanism for auxeticity in the case of the nanoplate is a special case of the rectangular nanowire.

The above work illustrates that auxeticity can be caused by the surface induced asymmetric stresses in the two lateral directions of metal nanowires.  Following this discovery, Kim et al. studied the Poisson's ratio for bulk cubic materials along principal directions with a proper lateral loading.\cite{hoPSSB2016b}  Previous works have shown that most cubic materials are auxetic in the non-principal directions,\cite{MilsteinF1979prb,BaughmanRH1998nat} but the Poisson's ratio was reported to be positive along the principal directions. However, Kim's group found that auxeticity can occur along principal directions in cubic materials if stresses are applied to two lateral directions with different magnitudes, which essentially mimics the surface induced asymmetric lateral stresses in metal nanowires.

\begin{figure}[tb]
  \begin{center}
    \scalebox{1.0}[1.0]{\includegraphics[width=8cm]{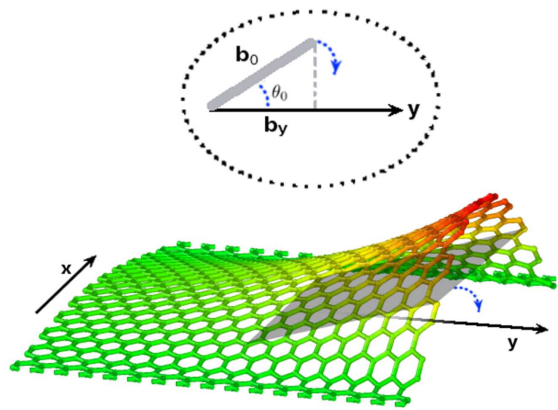}}
  \end{center}
  \caption{(Color online) Inclined plate model for warped edge induced NPR. The warped free edge is represented by the inclined plate (in gray). During the tensile deformation of graphene, the inclined plate falls down, which leads to the increase of its projection along the y-direction, resulting in the NPR effect. Reproduced with permission from Ref.~\onlinecite{JiangJW2016npr_fbc}. Copyright 2016, American Chemical Society.}
  \label{fig_ipmodel}
\end{figure}

\textbf{Warping free edge induced auxeticity for graphene.}  As previously discussed, there exist compressive edge stresses at the free edges of graphene, due to the under-coordinated edge atoms.\cite{ShenoyVB} Furthermore, graphene has a very large Young's modulus ($E$) but extremely small bending modulus ($D$), so it can be easily buckled upon compression according to the formula of the critical buckling strain,\cite{TimoshenkoS1987} $\epsilon_c=-4\pi^2 D/(EL^2)$. Consequently, graphene's free edges are buckled into a three-dimensional warping structure due to the compressive edge stresses. The warping structure can be well described by the surface function $z(x,y)=Ae^{-y/l_c}\sin (\pi x/\lambda)$, with $l_c$ as the penetration depth and $\lambda$ as the half wave length as shown in Fig.~\ref{fig_ipmodel}. 

Recently, two of the present authors showed that warping free edges can cause auxeticity for graphene ribbons with width less than 10~nm.\cite{JiangJW2016npr_fbc} It was found that the Poisson's ratio depends on the width of the ribbon and the magnitude of the applied strain. The Poisson's ratio stays negative for tensile strains smaller than about 0.5\%, and becomes positive when the applied strain is larger than this critical value. The critical strain corresponds to the structural transition of the edge from the three-dimensional warping configuration into the two-dimensional planar structure.

From an analytic point of view, each warping segment along the edge can be represented by an inclined plate, which falls down into the graphene plane during the stretching of the structure in the x-direction, leading to the increase of the projection of the inclined plate along the y-direction, which results in a negative value for the in-plane Poisson's ratio $\nu_{xy}$. Based on this inclined plate model, a general analytic expression was obtained for the Poisson's ratio in graphene ribbons of arbitrary width,
\begin{eqnarray}
\nu = \nu_{0}-\frac{2}{\tilde{W}}\left(\nu_{0}+\frac{1}{\epsilon_{c}}\tilde{A}^{2}C_{0}^{2}\right),
\label{eq_nu_ip_general}
\end{eqnarray}
where $C_{0}=\frac{2}{\pi}\left(1-\frac{1}{e}\right)$ is a constant and $\nu_{0}=0.34$ is the Poisson's ratio for bulk graphene. The critical strain is $\epsilon_c$, at which the warping structure becomes planar. The dimensionless quantity $\tilde{W}=W/l_{c}$ is the width with reference to the penetration depth $l_{c}$. The other dimensionless quantity $\tilde{A}=A/l_{c}$ is the warping amplitude with reference to the penetration depth. Eq.~(\ref{eq_nu_ip_general}) agrees quantitatively with the numerical simulation results.

\subsection{Extrinsic}

In contrast to intrinsic mechanisms, many nanomaterials exhibit NPR when their intrinsic structure and geometry, i.e. the flatness of 2D materials, is altered in some fashion due to external stimuli. For example, many of the examples of extrinsic NPR discussed below originate due to structural modifications that result in in-plane compression and out-of-plane deformation of the nanomaterials, particularly for 2D nanomaterials, such that when tensile strain is applied a flattening of the sheet resulting in a relative in-plane area expansion, and thus NPR, occurs.  We now discuss these extrinsic mechanisms that lead to NPR in nanomaterials. We note that these extrinsic mechanisms have also been used to induce NPR for bulk, or macroscale materials.

\subsubsection{Patterning}

\begin{figure}[tb]
  \begin{center}
    \scalebox{1.0}[1.0]{\includegraphics[width=8cm]{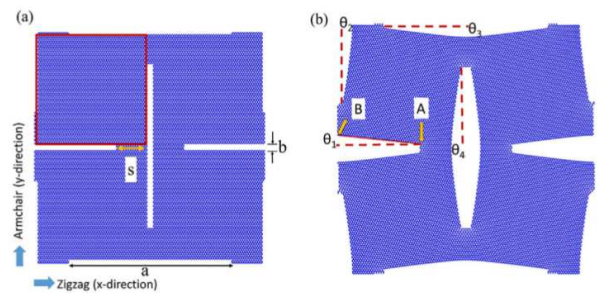}}
  \end{center}
  \caption{Patterned porous graphene under uniaxial loading. Size = $24.6\times 24.7$~{nm$^2$}. Configurations of the patterned porous graphene at strain (a) $\epsilon=0$ and (b) $\epsilon=0.1$ under uniaxial loading in the zigzag direction. Reproduced with permission.\cite{hoPSSB2016a} Copyright 2016, Wiley-VCH.}
  \label{fig_2016pssb_fig1}
\end{figure}

Patterning is a change of the material's structure, which has been widely employed to induce NPR in bulk materials.\cite{GrimaJN2000jmsl,SmithCW2000am,GasparN2005am,GrimaJN2006jms,RaviralaN2007jmsci}  For example, the notion of rotating squares were proposed to be a geometrical auxetic model by Grima and Evan in 2000.\cite{GrimaJN2000jmsl} Recently, one of the present authors and his collaborators introduced periodic cuts in graphene to mimic the rotating square model and obtained auxeticity in monolayer graphene for this particular structural pattern.\cite{hoPSSB2016a} Fig.~\ref{fig_2016pssb_fig1}~(a) shows the unit cell of the patterned graphene in which rectangular voids with the size $a\times b$ are perpendicularly distributed in graphene. The rotating units (in red rectangles) rotate to accommodate the applied strain. Except for the regions around the ends of the voids, the actual strain inside the graphene sheet is negligible.

As a result of the rotating effect, graphene expands in the lateral direction when it is stretched in one longitudinal direction, so the patterned graphene structure is auxetic. The ideal geometrical auxetic model has an isotropic Poisson's ratio of -1. Here, by changing the aspect ratio $a/b$ of the rectangular void, the Poisson's ratio can be tailored, and it approaches -1 as the void aspect ratio increases.  However, one problem of such patterned structures is failure at the two ends of the voids due to stress concentration. The failure might occur under large strains of 2\% for the case of patterned macroscopic metals.\cite{TaylorM2014am} However, because graphene can exhibit large elastic strains, the patterned graphene can exhibit NPR even at a strain of 20\%, and thus the rotating unit can be used in applications that require large deformation during operation.

\subsubsection{Buckling}

\textbf{Borophene.} In 2015, crystalline two-dimensional and atomically-thin boron sheets (borophene) were grown on silver surfaces under ultrahigh-vacuum conditions.\cite{MannixAJ2015sci} The low-energy borophene has the Pmmn space group symmetry with a rectangular unit cell of lattice constants $a=5.1\pm 0.2$~{\AA} and $b=2.9\pm 0.2$~{\AA}. A distinct feature for the borophene is the out-of-plane buckling along the \textbf{b} direction, in which the buckling height is about 0.8~{\AA}. The buckling configuration leads to strong anisotropic physical properties for the borophene. First-principles calculations predict that both in-plane Poisson's ratios are negative (-0.04 along \textbf{a} and -0.02 along \textbf{b}) due to the out-of-plane buckling.

A more recent first-principles calculation showed that borophene is also auxetic in the out-of-plane direction during stretching along the \textbf{b} direction.\cite{wangH2016njp} The out-of-plane auxeticity is mainly due to the weakening of the out-of-plane B-B bonding in stretched borophene, which is caused by flattening of the buckling height due to strong $\sigma$ bonds along the \textbf{a} direction.

\textbf{Penta-graphene.} Penta-graphene is a quasi-two-dimensional metastable carbon allotrope recently proposed by Zhang et al.\cite{ZhangS2015pnas} There are sp$^2$ and sp$^3$ hybridized carbon atoms in penta-graphene, resulting in a buckling height of 0.6~{\AA} in the out-of-plane direction. It was found that the in-plane Poisson's ratio is -0.068 for penta-graphene.\cite{ZhangS2015pnas} Although not elucidated in the original work of Zhang et al., this auxetic phenomenon is closely related to the finite buckling height in the out-of-plane direction.

\textbf{Graphane.} There have also been some first-principles calculations on the Poisson's ratio of fully hydrogenated graphene (graphane).\cite{CadelanoE2010prb,TopsakalM2010apl,PengQ2013pccp,AnsariR2015ssc} A finite buckling height of 0.65~{\AA} was obtained in the out-of-plane direction in the anisotropic boat-like fully hydrogenated graphene,\cite{CadelanoE2010prb} and first-principles calculations found that the Poisson's ratio can be negative in boat-like fully hydrogenated graphene.\cite{CadelanoE2010prb,ColomboL2011ppp}

\subsubsection{Rippling}

Atomically-thick 2D nanomaterials like graphene cannot exist as strictly planar crystals, because the Peierls transition will lead to considerable out-of-plane rippling at any finite temperature.\cite{PeierlsRE1935,LandauLD1937,fasolinoA2007nm}  Furthermore, the bending moduli for 2D nanomaterials like graphene\cite{arroyoM2004,luQ2009} or MoS$_{2}$\cite{JiangJW2013bend} are very small compared to the in-plane stiffness, so ripples can also be easily generated by external disturbance other than thermal fluctuations.

When a rippled structure is stretched, de-wrinkling and unfolding occurs, causing the flattening of the rippled conformation, and resulting in an expansion of the in-plane dimensions. Hence, the rippled structure exhibits auxeticity.  Below, we discuss methods by which ripples in graphene have been generated.

\textbf{Thermally-induced ripples.} Monte Carlo simulations show that the Poisson's ratio of graphene decreases with an increase in temperature, because the thermally-induced ripple amplitude is larger at higher temperature.\cite{ZakharchenkoKV} The Poisson's ratio can be negative at high temperatures, e.g. $-0.07\pm 0.18$ at 1700~K, and $-0.07\pm 0.21$ at 2100~K, and the reduction of the Poisson's ratio by increasing temperature was discussed from an entropic point of view. The structure tries to expand in the unstretched direction, so that the entropic energy can be minimized, resulting in the reduction of the Poisson's ratio.

However, it appears as though thermal vibrations are not an efficient approach to manipulate the Poisson's ratio, as very high temperature is needed to drive the Poisson's ratio into the negative regime. The amplitude of the thermal fluctuations depends on the bending modulus of graphene, so the temperature effect on the Poisson's ratio can be modified through changing the bending modulus value. Using the triangular mesh model, it was shown that increasing the bending modulus suppresses the thermal-induced ripples, leading to the weakening  of the auxeticity.\cite{UlissiZW2016acsn}

\textbf{Vacancy induced ripples.} We have discussed above that thermally-induced ripples are not an optimally efficient mechanism for tuning the auxeticity in graphene. In contrast, topological defects, which induce substantial local curvature, can cause significant amounts of rippling which is critical for manipulation of the Poisson's ratio in graphene.\cite{GrimaJN2015adm}

It was found that the 5-8-5 double vacancy defect is effective in inducing rippling (as compared with the thermally-induced ripples for pure graphene), with an increase in the number of ripples with increasing double vacancy defect density.  Grima et al. found that the Poisson's ratio can be negative (with -0.3 as the most negative value) for the graphene containing specific densities of 5-8-5 double vacancies at room temperature for strains less than a critical value.\cite{GrimaJN2015adm} The critical strain becomes larger for graphene with more defects.

\begin{figure}[tb]
  \begin{center}
    \scalebox{1}[1]{\includegraphics[width=8cm]{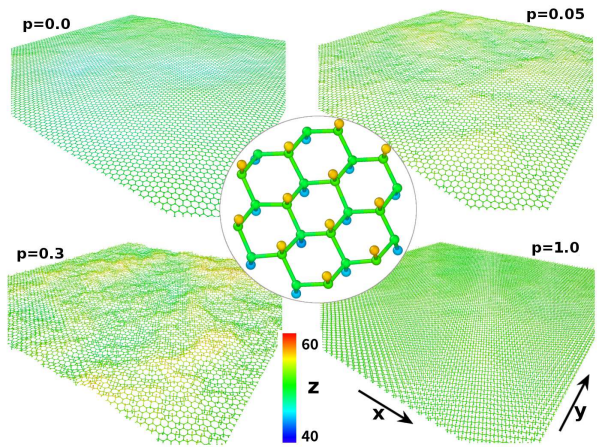}}
  \end{center}
  \caption{(Color online) Structure for hydrogenated graphene of size $200\times 200$~{\AA} at room temperature. The left top panel shows the pure graphene (i.e., percentage of hydrogenation $p=0.0$). The other three panels show the hydrogenated graphene with $p=0.05$, 0.3, and 1.0. The central inset shows the chairlike hydrogenation pattern for the fully hydrogenated graphene with $p=1.0$, where hydrogen atoms are bonded to carbon atoms on both sides of the plane in an alternating manner. The colorbar is with \rev{respect} to the z-coordinate of each atom. Reproduced from Ref.~\onlinecite{JiangJW2016npr_hydrogen} with permission from The Royal Society of Chemistry.}
  \label{fig_cfg_hydrogen}
\end{figure}

\begin{figure}[tb]
  \begin{center}
    \scalebox{1}[1]{\includegraphics[width=8cm]{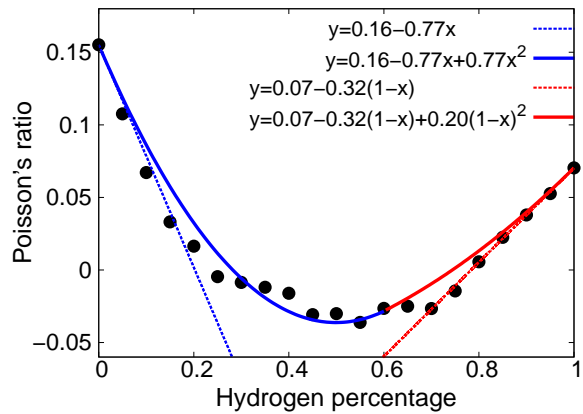}}
  \end{center}
  \caption{(Color online) The Poisson's ratio versus the percentage of hydrogenation for graphene of size $200\times 200$~{\AA} at room temperature. Reproduced from Ref.~\onlinecite{JiangJW2016npr_hydrogen} with permission from The Royal Society of Chemistry.}
  \label{fig_poisson_hydrogen}
\end{figure}

\textbf{Hydrogenation induced ripples.} One of the present authors and collaborators found that hydrogenation-induced ripples can effectively tune the Poisson's ratio from positive to negative for graphene.\cite{JiangJW2016npr_hydrogen} Fig.~\ref{fig_cfg_hydrogen} shows obvious ripples in the randomly hydrogenated graphene. The ripples have the largest amplitude for hydrogenated graphene with hydrogenation percentage around 50\%; i.e., the ripple amplitude becomes larger with increasing hydrogenation percentage below 50\%, and will become weaker with further increasing hydrogenation percentage. This is because both pure graphene and the fully hydrogenated graphene are perfect periodic crystals, which should not have the doping (hydrogenation) induced ripples. It is thus reasonable to have the largest amplitude for ripples in the partially hydrogenated graphene with some moderate hydrogenation percentages.

The Poisson's ratio for the hydrogenated graphene was found to be dependent on the hydrogenation percentage, as shown in Fig.~\ref{fig_poisson_hydrogen}. In particular, by increasing hydrogenation percentage, the Poisson's ratio reaches a minimum and negative value for half-hydrogenated graphene, which possesses the ripples with the largest amplitude. The ripples are weak in the hydrogenated graphene with hydrogenation percentages around 0 and 100\%, so the Poisson's ratio changes as a linear function of the hydrogenation percentage. However, the change of the Poisson's ratio becomes nonlinear in the highly rippled graphene with hydrogenation percentage around 50\%, because of correlations between neighboring ripples with large amplitude.

\begin{figure}[tb]
  \begin{center}
    \scalebox{1}[1]{\includegraphics[width=8cm]{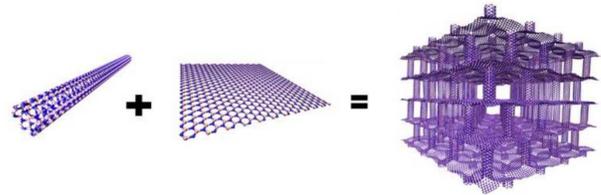}}
  \end{center}
  \caption{(Color online) Schematic picture of pillared boron nitride made of 1D boron nitride nanotube and 2D monolayer h-boron nitride sheets. Reproduced with permission from Ref.~\onlinecite{SakhavandN2014jpcc}. Copyright 2014, American Chemical Society.}
  \label{fig_2014jpcc_fig1}
\end{figure}

\textbf{Junction induced ripples.} In addition to the techniques discussed above for generating ripples in 2D materials, ripples can also be obtained in 3D porous structures. For example, in 2014, Sakhavand and Shahsavari introduced 3D boron nitride, namely, pillared boron nitride nanostructures in which parallel monolayers of h-boron nitride are connected by vertical single-walled boron nitride nanotubes (Fig.~\ref{fig_2014jpcc_fig1}).\cite{SakhavandN2014jpcc} The combination of the nanotubes and the monolayers causes large ripples at the junctions as well as in the sheets. As a result, the pillared boron nitride nanostructures can exhibit an in-plane NPR of -0.24 to -0.28. Similarly, pillared graphene nanostructures with the similar geometry can also exhibit a NPR of -0.10 to -0.14.\cite{SihnS2012carbon}

\subsubsection{Other mechanisms}

\begin{figure}[tb]
  \begin{center}
    \scalebox{1}[1]{\includegraphics[width=8cm]{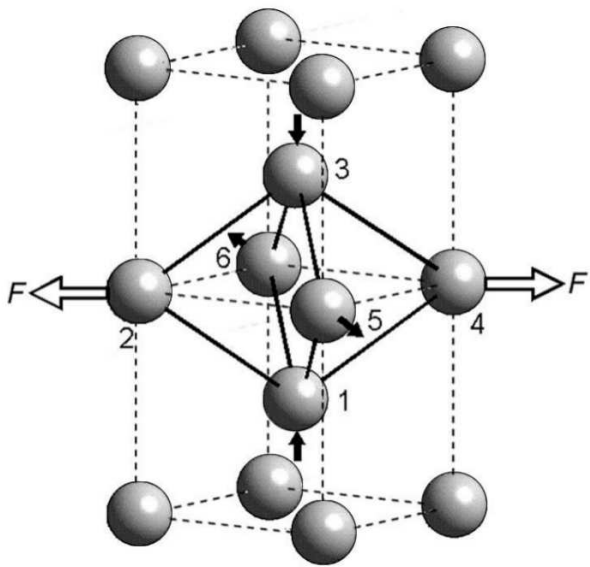}}
  \end{center}
  \caption{(Color online) The structural origin of a negative Poisson's ratio along the [$1\bar{1}0$]-direction as stretched along [110]-direction for the case of a rigid-sphere BCC crystal. The white arrows indicate the loading direction. Reprinted by permission from Macmillan Publishers Ltd: Nature\cite{BaughmanRH1998nat}, copyright 1998.}
  \label{fig_1998nat_fig3}
\end{figure}

\begin{figure}[tb]
  \begin{center}
    \scalebox{1}[1]{\includegraphics[width=8cm]{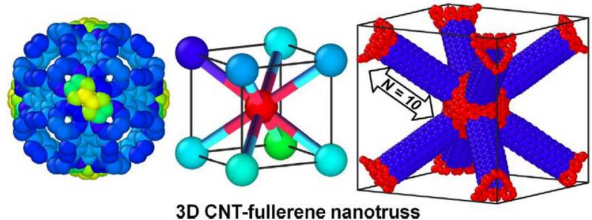}}
  \end{center}
  \caption{(Color online) Schematic illustration for the 3D carbon nanotube-fullerene nanotruss. A fullerene (left); a unit cell of BCC-lattice (middle); one representative unit cell of 3D nano-truss matrix (right), similar to a BCC unit cell. Reproduced with permission.\cite{WuJY2013cms} Copyright 2013, Elsevier.}
  \label{fig_2013cms}
\end{figure}

\textbf{Cubic crystals.} Some crystals exhibit NPR if they are loaded in a particular direction at both nano and macro length scales.\cite{LethbridgeZAD2010am,MilsteinF,BaughmanRH1998nat} As mentioned above, 69\% of the cubic elemental metals show NPR if stretched along the [110]-direction.\cite{BaughmanRH1998nat} Fig.~\ref{fig_1998nat_fig3} provides simple geometrical arguments based on the pairwise central force assumption to explain the auxetic mechanism for a BCC crystal. Under tensile loading along the [110]-direction (marked by the white arrows), decreasing the angle 143 is the only way to maintain the length of the bonds, resulting in the decrease of the distance between atoms 1 and 3 along the [100]-direction. The movement of atoms 1 and 3 causes atoms 5 and 6 to move apart along the [$1\bar{1}0$]-direction, providing a negative Poisson's ratio.\cite{BaughmanRH1998nat} Based on this simple analysis, the atoms move rigidly and the value of the NPR is -1.

Wu et al. proposed a 3D nano-truss matrix in which (6,6) carbon nanotubes with length N are connected by coalescence of fullerenes, where the similarity of the truss structure to a BCC unit cell (Fig.~\ref{fig_2013cms}) explains the resulting auxeticity.\cite{WuJY2013cms} The Poisson's ratio $\nu$ is strongly dependent on the length of the carbon nanotube. As the length increases, the rigid movement mechanism of the fullerenes is more significant and thus the auxeticity is higher.

\textbf{Saint-Venant effects.} As mentioned above, pristine single-walled carbon nanotube might theoretically exhibit auxeticity, although it is difficult to realize in practice. However, the Poisson's ratio of single-walled carbon nanotubes with non-reconstructed vacancies (i.e. ideal vacancy defects at 0~K) show large variation and is dependent on the tube geometry, the percentage, and location of vacancies.\cite{ScarpaF2009jpdap} When the defects are close to the end of the nanotube, the Saint-Venant effects on the loading condition, i.e., the non-uniform distribution of the applied axial forces around the vacancies, is significant. \rev{We note that the Saint Venant's principle refers to the notion that stress from self equilibrated load distributions tend to decay much more rapidly with distance than stress or strain from distributions equivalent to a net force or moment.} As a result, there is out-of-plane rotation of C-C bonds that is connected to the vacancies and local radial expansion of the nanotube under tension. Hence, locally NPR can be obtained in the nanotube.

\section{Experimental Studies on Nanomaterial NPR}
\subsection{Auxeticity for pure nanomaterials}
\subsubsection{Black phosphorus}

\begin{figure}[tb]
  \begin{center}
    \scalebox{0.8}[0.8]{\includegraphics[width=8cm]{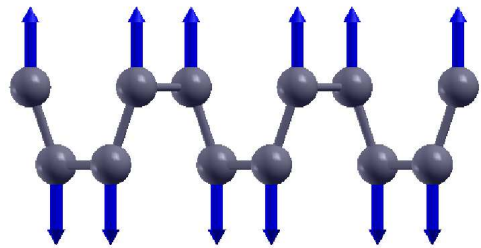}}
  \end{center}
  \caption{Vibration morphology for the~{A$_{g}^{1}$} phonon mode in black phosphorus.}
  \label{fig_bp_A1g_mode}
\end{figure}

For bulk materials, the Poisson's ratio can be measured directly by recording the position of location markers during the loading process, while the out-of-plane Poisson's ratio can be obtained with the thickness determined by scanning electron microscopy (SEM).\cite{GasparN2005am}  In contrast, very few experimental studies of the auxetic behavior of nanomaterials have been performed.  However, for nanoscale 2D black phosphorous, a recent experiment provided indirect evidence for the auxeticity of black phosphorus by measuring the strain-induced frequency shift of the~{A$_{g}^{1}$} phonon mode displayed in Fig.~\ref{fig_bp_A1g_mode}.\cite{DuY2016arxiv}

The vibrational frequency for the~{A$_{g}^{1}$} phonon mode is shifted by uniaxial in-plane strains for black phosphorus. The frequency is reduced when black phosphorus is stretched along the armchair or zigzag in-plane direction. This frequency shift has also been measured in previous experiments.\cite{WangY2015nnr,LiY2016afm} From the vibration morphology of the~{A$_{g}^{1}$} mode, the frequency of this phonon mode is related to the inter-group bond in the out-of-plane direction. For few-layer (or bulk) black phosphorus, its frequency is also related to the space between adjacent black phosphorus layers. The experimentally-observed reduction of the frequency for this phonon mode indicates that either the inter-group bond length or the inter-layer space has been enlarged, as a larger inter-atomic distance leads to a weaker atomic interaction for the P-P bonds.\cite{DuY2016arxiv} Hence, the stretching induced reduction of the frequency for the phonon mode provides evidence for the auxeticity of black phosphorous that has been obtained theoretically.\cite{JiangJW2014bpnpr,ElahiM2014prb,QinGarxiv14060261}

\subsection{Auxeticity for nanomaterial composites}

While experiments for auxetic behavior in individual nanomaterials have been rare, there have been some experiments demonstrating auxeticity for composites containing nanomaterials. These composites are typically on the macroscopic size scale, enabling the Poisson's ratio to be measured using the standard photograph technique or by recording the loading induced structural deformation with the help of location markers.

\subsubsection{Carbon nanotube Sheets and Films}

\begin{figure}[tb]
  \begin{center}
    \scalebox{0.8}[0.8]{\includegraphics[width=8cm]{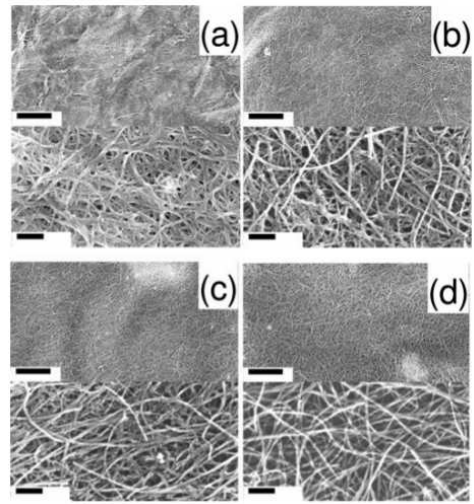}}
  \end{center}
  \caption{SEM images of the surfaces of buckypaper containing (a) 0 wt.\%, (b) 47.1 wt.\%, (c) 72.7 wt.\%, and (d) 100 wt.\% multi-walled carbon nanotube content. Different magnifications are shown in top and bottom parts of each image. The scale bars for the upper and lower images in (a)-(d) correspond to 2~{$\mu$m} and 200~{nm}, respectively. Reprinted figure with permission from V. R. Coluci et al. Physical Review B 78, 115408 (2008).\cite{ColuciVR2008prb} Copyright 2008 by American Physical Society.}
  \label{fig_2008buckypaper}
\end{figure}

While the Poisson's ratio for both single-walled and multi-walled carbon nanotubes is positive, it was found that the Poisson's ratio for carbon nanotube sheets containing fiber networks (buckypaper) can be reduced by increasing the weight percentage of multi-walled carbon nanotubes in the composites.\cite{HallLJ2008sci,ColuciVR2008prb} The in-plane Poisson's ratio for the nanotube sheets becomes negative if the weight percentage of multi-walled carbon nanotube exceeds about 73\%, where a limiting NPR value of -0.2 can be achieved in nanotube sheets with 100 weight percent of multi-walled carbon nanotubes. Considering that both single- and multi-walled carbon nanotubes have positive Poisson's ratio, the auxeticity for nanotube sheets should result from the assembly of nanotubes in the complex sheet network shown in Fig.~\ref{fig_2008buckypaper}.

In the measurement of the in-plane Poisson's ratio, nanotube sheets were coated with trace amounts of TiO$_2$ particles for position marking. Digital images captured for the nanotube sheets were analyzed using image correlation software to determine the variation of the distance between TiO$_2$, giving the lateral strains during the deformation process. To measure the out-of-plane Poisson's ratio, the change of the thickness was obtained by SEM during the deformation process.

While nanotubes in this experiment are randomly assembled, another experiment demonstrated that the in-plane Poisson's ratio can be decreased to -0.5 for sheets with highly oriented carbon nanotubes.\cite{ChenL2009apl} The auxetic property can be maintained (with value -0.53) by embedding carbon nanotubes in the polymer matrix. Their theoretical model indicates that the auxeticity can be attributed to the realignment of curved nanotubes during stretching.

\begin{figure}[tb]
  \begin{center}
    \scalebox{0.8}[0.8]{\includegraphics[width=8cm]{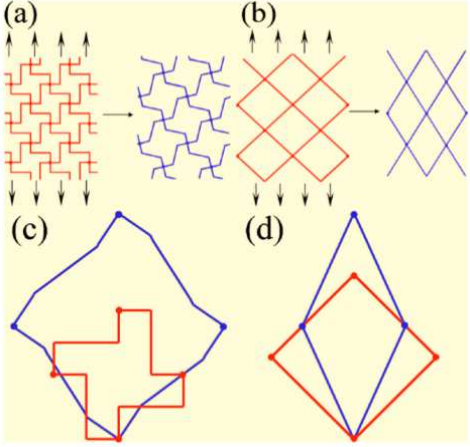}}
  \end{center}
  \caption{Deformation mechanisms [(a) and (b)] of carbon nanotube films for negative and positive Poisson's ratios. [(c) and (d)] The evolutions of the basic unit structure for negative and positive Poisson's ratio. Reproduced with permission.\cite{MaYJ2010apl} Copyright 2010, AIP Publishing.}
  \label{fig_2010apl}
\end{figure}

In another experiment, it was reported that the sign of the Poisson's ratio for the carbon nanotube film depends on the strain during stretching.\cite{MaYJ2010apl} The Poisson's ratio changes from negative to positive when the applied tensile strain is larger than a critical value. The critical strain value is larger in the nanotube film constructed by thinner carbon nanotubes, which is explained using the theoretical model shown in Fig.~\ref{fig_2010apl}. Carbon nanotubes are rippled in the films. For smaller strain, the deformation of the nanotube film can be modeled by Fig.~\ref{fig_2010apl}~(a) and (c), where curved carbon nanotubes are straightened, resulting in the expansion of the film during stretching, and thus a NPR.  When most carbon nanotubes have been straightened, a further stretching of the nanotube film can be simulated by Fig.~\ref{fig_2010apl}~(b) and (d), which yields a positive value for the Poisson's ratio. For films with thinner nanotubes, the critical strain is larger, because more strain is needed to straighten all carbon nanotubes, considering that thinner nanotubes undergo more bending in Fig.~\ref{fig_2010apl}~(a).

\subsubsection{Graphene metamaterials}

Recently, Zhang et al. observed auxeticity in a three-dimensional graphene metamaterial with ordered hyperbolic pattern and hierarchical honeycomb-like scaffold of microstructure.\cite{ZhangQ2016am} A Poisson's ratio value of -0.38 was observed in a properly designed sample with local oriented `buckling' of multilayer graphene cellular walls as the microstructure. The evolution of the microscopic structure during compression was  monitored using SEM.

The auxeticity was attributed to the microstructure of orthogonal-hyperbolic pattern, which is realized by proper freeze-casting orientation and macroscopic aspect ratio. In this pattern, graphene based cellular walls will present oriented buckling-induced ripples during compression, which provides the fundamental mechanism for the auxeticity of the graphene metamaterial. We note that this experiment sheds light on the ripple-induced auxeticity mechanism in graphene as discussed in the previous theoretical section, as the subsequent freezing process (closely related to ripples) plays an important role in manipulating the Poisson's ratio in this experiment.

\section{Future prospects and summary}

\subsection{More Experimental Studies Needed}

From the above, it is clear that more experiments are needed for the field of auxeticity in nanomaterials. The Poisson's ratio for bulk materials can be measured by directly recording the structural evolution during deformation, such as taking photographs at the macroscale or using the SEM at the microscale.\cite{LakesRS1987sci,FriisEA1988jms,CaddockBD1989jpdap,ColuciVR2008prb,ChenL2009apl} The lateral resultant strain and the applied strain can be measured simultaneously by analyzing the photographs or the SEM pictures using digital correlation analysis software. For example, in carbon nanotube sheets, the in-plane Poisson's ratio is measured by taking photographs while the out-of-plane Poisson's ratio is obtained from the \rev{SEM images}.\cite{ColuciVR2008prb} Sometimes, markers will be introduced to facilitate a more accurate recording for the structural deformation of the specimen during loading test.\cite{CaddockBD1989jpdap,ColuciVR2008prb}

For low-dimensional nanomaterials, we are only aware of one indirect detection of the auxetic phenomenon in black phosphorus,\cite{DuY2016arxiv} while direct experiments are still lacking. However, it may be possible to use direct imaging techniques such as the SEM to measure the Poisson's ratio considering recent demonstrations of experimental capability in manipulating and deforming graphene to very large ($>$200\%) uniaxial strains. Specifically, Blees et al. were able to capture the structure change during the stretching of the graphene kirigami (`kiru', cut; `kami', paper) using the SEM technique,\cite{BleesMK2015nat} which was also predicted through MD simulations.\cite{qiPRB2014} Within this experimental setup, it may be possible to directly measure the Poisson's ratio for pure graphene or specifically engineered graphene, for which the auxeticity has been theoretically predicted.

In addition, many of the specifically engineered structures are also realizable in current experiments.  The hydrogenation process can be realized experimentally and the process is reversible.\cite{SofoJO2007prb,EliasDC2009sci} Experiments have demonstrated a fairly good degree of control over the vacancy defects in graphene.\cite{EsquinaziP2003prl,HanKH2003am} The thermally-induced ripples can be manipulated utilizing the difference in the thermal expansion coefficient of graphene and substrate.\cite{BaoW2009nn} According to very recent experiments,\cite{ZhangQ2016am} the thermally-induced ripple is a possible mechanism for the auxeticity observed in graphene metamaterials.  Furthermore, nanomaterials can usually sustain large mechanical strain, and the strain can be engineered over a wide range in these nanomaterials, which eases the measurement of the Poisson's ratio.\cite{NiZH2008acsn,MohiuddinTMG2009prb,GarzaHHP2014nl} 

\subsection{More auxetic nanomaterials}

\subsubsection{Search for auxetic nanomaterials}

There are several directions that can be pursued for exploring the mechanisms underpinning auxeticity, or applications exploiting the newly uncovered property of auxeticity in nanomaterials.

\textbf{New auxetic mechanisms.} Nanomaterials have some novel features as compared with bulk materials, and some of these novel features can induce considerable auxeticity. For example, surface and edge effects can dominate mechanical properties in nanomaterials, due to their large surface to volume ratios. The out-of-plane ripples in two-dimensional atomic-thick nanomaterials can induce strong effects on various physical properties.  Therefore, it is important to examine additional out-of-plane deformation mechanisms similar to edge or rippling effects that are unique to nanomaterials.  For example, besides the common Poisson's ratio related to tension, it may be also interesting to search for possible auxeticity in the bending Poisson's ratio.\cite{LiuX2014prl}

\textbf{More auxetic nanomaterials.} Auxeticity has been shown to be intrinsic for black phosphorus, graphene and borophene. A natural question arises: will the auxeticity be an intrinsic property for other nanomaterials? Hence, it is necessary to examine possible auxetic phenomenon for h-BN, MoS$_2$ and etc. In particular, the thermal or defect induced rippling is a characteristic feature for all two-dimensional nanomaterials, so the ripple-induced auxeticity may  occur in other two-dimensional nanomaterials beyond graphene. The auxeticity resulting from the competition between multiple deformation modes\cite{JiangJW2016npr_intrinsic} may also be found in these two-dimensional nanomaterials, which exhibit similar competition between the bond stretching and angle bending interactions.

\textbf{Geometrically patterned auxetic nanomaterials} Graphene can be cut into specific geometric patterns, which can be regarded as nanoscale counterparts for previous geometrical auxetic models that were applied to bulk materials and structures. It will be of practical significance to examine possible auxeticity in graphene and other two-dimensional nanomaterials that are cut following specific auxetic geometries.  For example, the rotating squares model was proposed by Grima and Evans as an auxetic mechanism,\cite{GrimaJN2000jmsl} and this geometrical model is realized by Kim's group using patterned graphene.\cite{hoPSSB2016a} As another example, some periodic pleated origami geometries are predicted to be auxetic,\cite{WeiZY2013prl} which should be able to be realized in the experiment using the present graphene-based kirigami technique.\cite{BleesMK2015nat} Finally, it will also be important to determine if new patterns may emerge at the nanoscale by exploiting certain physics, i.e. edge and surface effects, ultralow bending moduli, etc, that are only observed in nanomaterials.

\textbf{Auxetic nanomaterial composites.} While it is certainly more challenging to measure Poisson's ratio at the nanoscale, the assembly of nanomaterials can result in macroscale structures, where the experimental equipment needed to measure the Poisson's ratio is 
well established. This may have important practical consequences, as many of the most interesting applications of nanomaterials occur at the macroscale.  For example, silicon nanowires are usually assembled into silicon nanowire networks for specific applications as electric devices\cite{MorelPH2012apl,TernonC2013pssrrl,SerreP2015nano,KelesU2013apl} or solar cells.\cite{OenerSZ2016nl} Furthermore, the assembly of nanomaterials will result in many new mechanisms for generating auxeticity due to the complex interactions between nanomaterials.  Researchers interested in this direction may draw initial inspiration from early successes such as auxetic carbon nanotube fibers\cite{HallLJ2008sci,ColuciVR2008prb,ChenL2009apl,MaYJ2010apl} and auxetic graphene metamaterials.\cite{ZhangQ2016am}

\subsubsection{Design of auxetic nanomaterials and nanostructures}

Besides searching for auxeticity in existing nanomaterials, it is also important to develop proper optimization approaches for the design of auxetic nanomaterials. Such studies have to-date been done on an ad hoc basis for nanomaterials, as a systematic approach for the design and optimization of auxetic nanomaterials has not been developed, which is in contrast to the large literature that has emerged regarding the design and optimization of bulk auxetic materials.\cite{TheocarisPS1998so,ZhangR1999jrpc,GuoX2010ams,SchwerdtfegerJ2011adm,MitschkeH2011adm,WangY2014cms,ZhouG2016smo}  In a similar way, it may be possible to design auxetic structures on the nanoscale level with some specifically chosen basic nanoscale elements such as nanoscale defects, edge/surface effects etc.\cite{LahiriJ2010enn,CarrLD2010nn} 

\subsection{Applications of auxetic nanomaterials}

\subsubsection{Novel applications}

While the search for more auxetic mechanisms and nanomaterials is essential, it is also important to begin investigating possible applications for auxetic nanomaterials.  Besides the applications commonly proposed for bulk auxetic materials, i.e. tougher materials,\cite{ChoiJB1992jms2} national security and defense, sound and vibration absorption,\cite{LipsettAW1988jasm,AldersonKL2000jms} it is likely that new, unexpected applications may emerge for nanoscale auxetics due to their unique, nanoscale dimensionality and properties.  For example, the lattice constants of some graphene allotropes are tunable, which enables the hollow lattice structure to be used as a filter.\cite{XueM2013nano} As one possible application, auxeticity can be exploited in cleaning filters by stretching, as the pore size will increase due to the NPR.  

Graphene with nanopores can be used to determine the sequence of DNA molecules with high resolution, which takes advantage of its high stiffness and high in-plane electrical conductivity.\cite{GarajS2010nat,MerchantCA2010nl,SchneiderGF2010nl,AhmedT2012nl,AvdoshenkoSM2013nl,HeeremaSJ2016nn} Theoretical studies have shown that these free edges can induce auxeticity in graphene for strain less than 1\%.\cite{JiangJW2016npr_fbc} Using the abnormal geometrical response (expand upon stretching) of the auxetic graphene hole, it may be possible to tune the contact between the graphene hole and the DNA molecules, so that the DNA sequencing process can be finely tuned through mechanical strain. These are two speculative applications of auxetic nanomaterials, though we expect that many more will emerge through the ingenious nature of the scientific community.

\subsubsection{Auxetic effects on physical properties}

A final, and likely intellectually rich area for future investigation from both the fundamental scientific point of view as well as the practical application point of view, is the consideration of the coupling of auxetic mechanical properties on other physical properties in nanomaterials.

For bulk materials, the auxetic effects on various physical properties have been widely examined.\cite{LipsettAW1988jasm,ChenCP1991jms,ChoiJB1992jms2,AldersonKL1994amm,ChoiJB1996ijf,ChanN1999jcp,ChanN1999jcp2,AldersonKL2000pes,ScarpaF2003aj,LiuB2006mm,SparavignaA2007prb,ShangXC2007pssb,SongF2008prl,LuZ2011pssb,MaT2012pbcm,ZhangZK2013sms,MaY2013sms,HeG2014sms,GoldsteinRV2014ms,LimTC2014pssb,HouS2015md,RenX2016sms,WangY2016sms} The auxetic effects can be obtained directly from the analytic formulas for isotropic materials within the classical elasticity theory. For example, in three-dimensional isotropic materials, the speed of sound is proportional to $(1+\nu)^{-1/2}$, and the hardness is related to $(1-\nu^2)^{\gamma}$ with $\gamma$ as a constant. Hence, auxeticity with $\nu\rightarrow -1$ leads to the enhanced speed of sound and enhanced toughness.

However, for nanomaterials, very few studies have examined the connection between the Poisson's ratio and other physical properties.\cite{HouJ2016ams} Furthermore, according to Eq.~(\ref{eq_nu_vffm}), the Poisson's ratio for graphene is in the range $-1/3 < \nu < 1$. \rev{In particular, the most negative value for the Poisson's ratio (-1/3 for graphene) is different from the lower bound of -1 for two or three-dimensional isotropic materials.  Furthermore, the lower bound value of -1/3 previously reported for graphene\cite{ChangT2003jmps} may be different for other atomically thin 2D materials.} This implies that traditional bulk auxetic applications, such as enhanced speeds of sound or toughness, may need to be re-examined for graphene and probably other nanomaterials.  Thus, it is necessary to develop an analytical and mechanistic understanding of how auxeticity impacts the other physical properties in nanomaterials.

\textbf{Acknowledgements} The authors thank Tienchong Chang for critical reading of the manuscript. The work is supported by the Recruitment Program of Global Youth Experts of China, the National Natural Science Foundation of China (NSFC) under Grant No. 11504225, and the start-up funding from Shanghai University. SYK acknowledges the support from the Mid-Career Researcher Support Program (grant no. 2014R1A2A2A09052374) of the National Research Foundation (NRF) of Korea. HSP acknowledges the support of the Mechanical Engineering department at Boston University.


\end{document}